\documentclass{emulateapj}
\usepackage{amsmath}
\usepackage{verbatim}
\usepackage[breaklinks,colorlinks,citecolor=blue,urlcolor=blue]{hyperref}
\usepackage[all]{hypcap}

\newcommand{\angstrom}{\mbox{\normalfont\AA}}
\newcommand{\tsc}[1]{~\textsc{#1}}
\newcommand{\nh}[1]{N(H$_2$)$_{\mathrm{#1}}$}

\shorttitle{ULTRAVIOLET \emph{HST} SPECTROSCOPY OF \emph{PLANCK} COLD CLUMPS}
\shortauthors{DIRKS \& MEYER}

\begin{document}


\title{Ultraviolet \emph{HST} Spectroscopy of \emph{Planck} Cold Clumps}


\author{Cody Dirks and David M. Meyer}
\affil{Center for Interdisciplinary Exploration and Research in Astrophysics, Dept. of Physics \& Astronomy,\\ Northwestern University, 2145 Sheridan Road, Evanston, IL, 60208;\\
codydirks2017@u.northwestern.edu}

\begin{abstract}
We report results of the first study utilizing the ultraviolet capabilities of the \emph{Hubble Space Telescope} to investigate a sample of \emph{Planck} Galactic Cold Clump (PGCC) sources. We have selected high-resolution spectra toward 25 stars that contain a multitude of interstellar absorption lines associated with the interstellar medium (ISM) gas within these PGCC sources, including carbon monoxide (CO), C\tsc{i} and O\tsc{i}. By building cloud-component models of the individual absorption components present in these spectra, we can identify and isolate components associated with the PGCC sources, allowing for a more accurate investigation of the ISM behavior within these sources. Despite probing a broad range of overall sightline properties, we detect CO along each sightline. Sightlines with CO column density N(CO)~$>$~10$^{15}$~cm$^{-2}$ exhibit spatial dependence in N(CO) and CO/C\tsc{i}, while sightlines with N(CO)~$<$~10$^{15}$~cm$^{-2}$ show no such spatial dependence. Differences between N(H$_2$) values derived from UV absorption and dust emission suggest structure in the spatial distribution of N(H$_2$), where ``CO-bright'' sightlines are associated with PGCC sources embedded within smooth translucent envelopes, and ``CO-dark'' sightlines are associated with PGCC sources embedded in patchier environments containing more diffuse gas.
\end{abstract}

\keywords{ISM: atoms -- ISM: clouds -- ISM: dust}

\section{Introduction}
Understanding star formation in the interstellar medium (ISM) entails unraveling the process by which the ISM transitions from the diffuse phase that fills the majority of the volume of the Galaxy to a dense, molecular phase where the majority of the mass and star formation resides. Due to the difficulty in directly observing the H$_2$ molecule (which constitutes the bulk of the mass in these regions) in radio frequencies, the molecular content of the ISM is frequently studied through observations of carbon monoxide (CO) emission, which has long been known to trace the dense, molecular phase of the ISM where star formation is likely to occur \citep[][and references therein]{Heyer2015}. As such, a wealth of studies have been performed at radio wavelengths investigating the observational relationship between CO and H$_2$ in these regions \citep[][and references therein]{Bolatto2013}. Further studies have taken place in the ultraviolet (UV) regime, due to the ability to directly measure the abundance of both CO and H$_2$ through absorption-line spectroscopy \citep{Burgh2007}. 

In an attempt to characterize the population of potential star-forming gas in the Milky Way, \citet{PlanckCollaboration2015} used dust emission maps to compile the \emph{Planck} Catalogue of Galactic Cold Clumps (PGCC), which represents an unbiased, all-sky census of 13,188 sources of cold dust emission. These sources have typical angular sizes of a few arcminutes, but due to the broad range of distances at which these sources are detected, their corresponding physical sizes range from $\sim$0.1 pc (for small, nearby cores) up to $\sim$30 pc (for distant, giant molecular cloud sources).  The gas content of many of these PGCC sources has been studied more carefully through follow-up observations (see \citet{Yuan2016,Liu2018}, as well as \citet{Wu2012}, which studied a precursor to the PGCC known as the Early Cold Core (ECC) Catalogue). These follow-up studies primarily observed CO at radio wavelengths, which is typically sensitive to column densities above N(CO)~$\sim$~10$^{15}$~cm$^{-2}$ \citep{Liszt2007}. Because of this CO detection limit, the vast majority of the work studying these clouds is biased toward the densest objects, rather than a representative sample of the entire PGCC population (Figure \ref{prev_pgcc_studies}) which contains a large number of sources not expected to show CO emission. These sources would be traditionally classified as diffuse or translucent based on their molecular hydrogen content, following the cloud classification scheme of \citet{Snow2006}.

The translucent regime has been studied in ultraviolet absorption by \citet{Burgh2010}, who studied the behavior of atomic and molecular carbon along 31 Galactic sightlines using the Space Telescope Imaging Spectrograph (STIS) aboard the  \emph{Hubble Space Telescope} (\emph{HST}). They determined that the carbon content of a particular sightline can determine the presence of translucent material; notably, they found that the CO/H$_2$ and CO/C\tsc{i} ratios were well correlated and could serve as a discriminator between diffuse and translucent material. They also found a sharp transition in the sightline-integrated N(CO), corresponding to when the cloud becomes dense enough for self-shielding to prevent photodissociation of CO.

To our knowledge, no study has yet been performed to specifically analyze PGCC sources in the UV regime. In this paper, we aim to apply the tools of UV absorption-line spectroscopy to sightlines selected to lie in the vicinity of PGCC sources, allowing us to study a subset of these sources that more accurately represents the entire PGCC population, which is expected to include many diffuse molecular sources which likely contain very little CO \citep{Snow2006}. Furthermore, in an effort to characterize these objects as accurately as possible, we seek to isolate the absorption components arising specifically from the PGCC source of interest. While this may introduce more uncertainty than simply integrating along the entire absorption profile (due to the potential complications involved in disentangling overlapping components), the effects of integrating the entire profile can skew key diagnostics of the ISM away from the true value in the local area of interest by convolving multiple unrelated areas of the ISM along the sightline.

In choosing a broader array of PGCC sources to sample and by isolating the relevant component along each sightline, we seek to better understand the PGCC population as a whole, and the role that these cold, dusty sources play in the formation of traditional molecular clouds within star-forming regions. This paper is organized as follows: Section 2 details our process for identifying our UV spectroscopic dataset, isolating gas associated with the PGCC source, and determining the relevant column densities. Section 3 details our results derived from the UV data, focusing on the behavior of the carbon and molecular hydrogen content along these sightlines. Section 4 discusses these results and describes a physical picture of the ISM along these sightlines sufficient consistent with the observed behavior. Section 5 summarizes our findings.

\section{Data Selection \& Reduction}
\subsection{Sightline Selection Process}
In order to isolate and study this transition regime, we seek to identify a sample of PGCC sources for which UV spectroscopy can be performed. This requires identifying stars whose sightlines lie in the sky vicinity of a PGCC source, as well as the ability to confidently place the star behind the gas of interest. To achieve this, we first use the sky locations and angular sizes of each of these sources to search the HST data archive for high-resolution (R $\geq$ 100,000) far-UV STIS spectra of bright OB stars in the vicinity of each source. In order to capture both the PGCC sources as well as the surrounding transition regions, we collect all sightlines that lie within 10$\sigma$ of the center of a nearby source, where $\sigma$ represents the angular size of an individual PGCC object (see Figure \ref{bullseye}).  
In order to ensure that we are probing the ISM gas associated with the PGCC source, we employ one of two methods. First, for PGCC sources with distances given in \citet{PlanckCollaboration2015}, we compared these source distances to stellar distances as given in the recently published Gaia Data Release 2 \citep{GaiaCollaboration2018} and identified cases where the minimum 1$\sigma$ error bound of the stellar distance is greater than the maximum 1$\sigma$ error bound of the source distance. This yielded 11 sightlines where the star can be placed behind the source. For PGCC sources with no distance estimate, or for sightlines where the errors in the stellar distance and the source distance are such that it is unclear whether or not the star is behind the source, we then compared CO emission from the Columbia-CfA Survey \citep{Dame2001} in the vicinity of the sightline to CO absorption observed in the UV spectra. CO emission studies are only sensitive to much larger column densities (N(CO)~$\gtrsim10^{15}$~cm$^{-2}$) than absorption (N(CO)~$\gtrsim10^{13}$~cm$^{-2}$), thus if the strongest observed absorption components occurred at the same velocity as those seen in emission, we can be certain that the absorption sightline is probing the gas associated with the PGCC source. This yielded a further 14 sightlines, giving a total of 25 suitable sightlines with \emph{HST} STIS spectra. These targets, their associated nearby PGCC sources, and the separation between the target and source are listed in Table \ref{results_table}.

\begin{figure}[]
\centering
\includegraphics[scale=0.24]{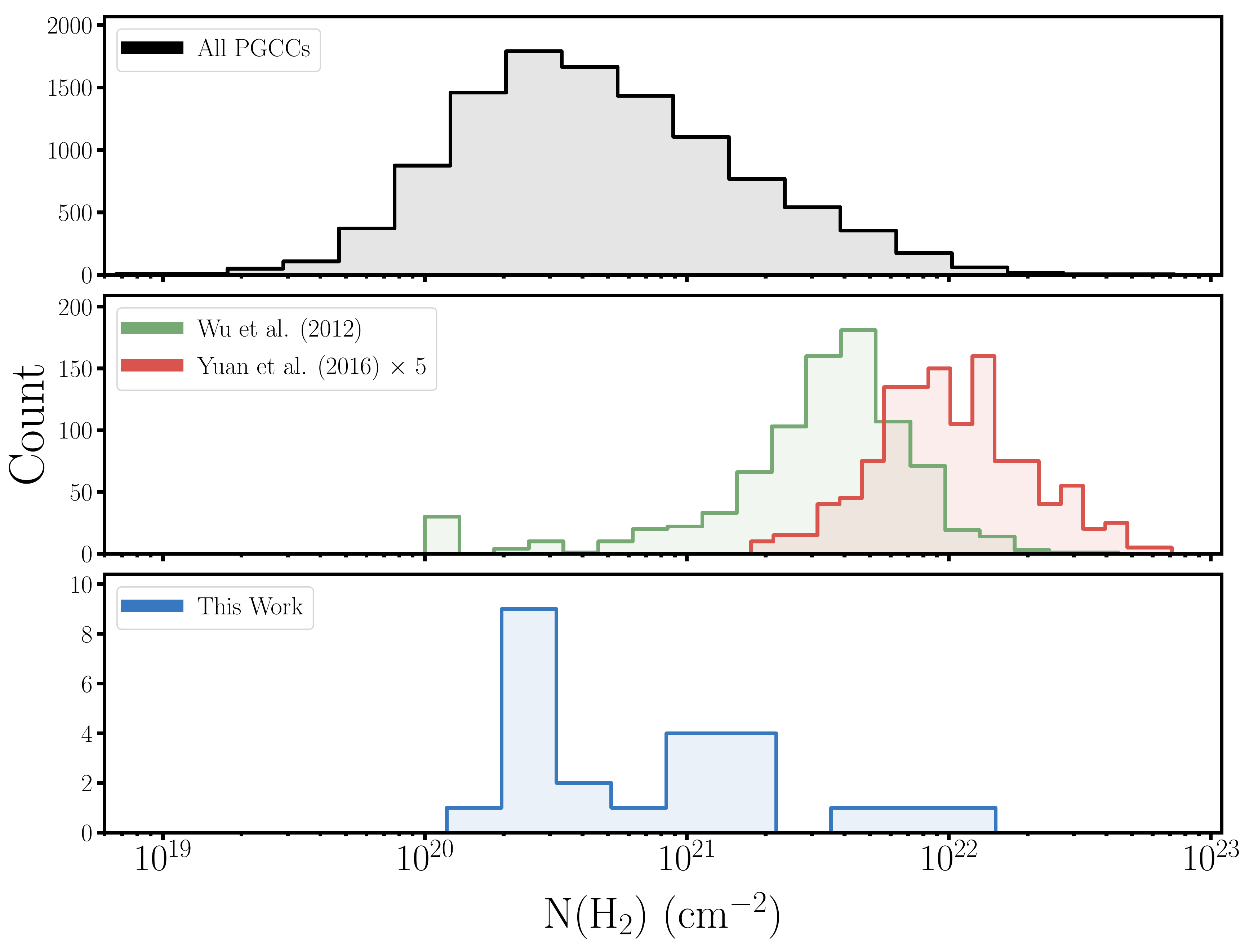}
\caption{Comparison of molecular hydrogen content of sampled PGCC sources from various studies. The entire catalogue \citep[top panel]{PlanckCollaboration2015} covers a broad range in N(H$_2$) and thus a broad range of ISM conditions. Radio emission studies (middle panel) primarily focus on the high N(H$_2$) end of this distribution, as these are the sources more likely to have large enough N(CO) to have detectable CO line emission. UV absorption techniques used in this work are more sensitive to diffuse molecular gas, and thus our sample probes a set of PGCC sources representative of the entire catalogue.}
\label{prev_pgcc_studies}
\end{figure}

\begin{figure}[]
\centering
\includegraphics[scale=0.24]{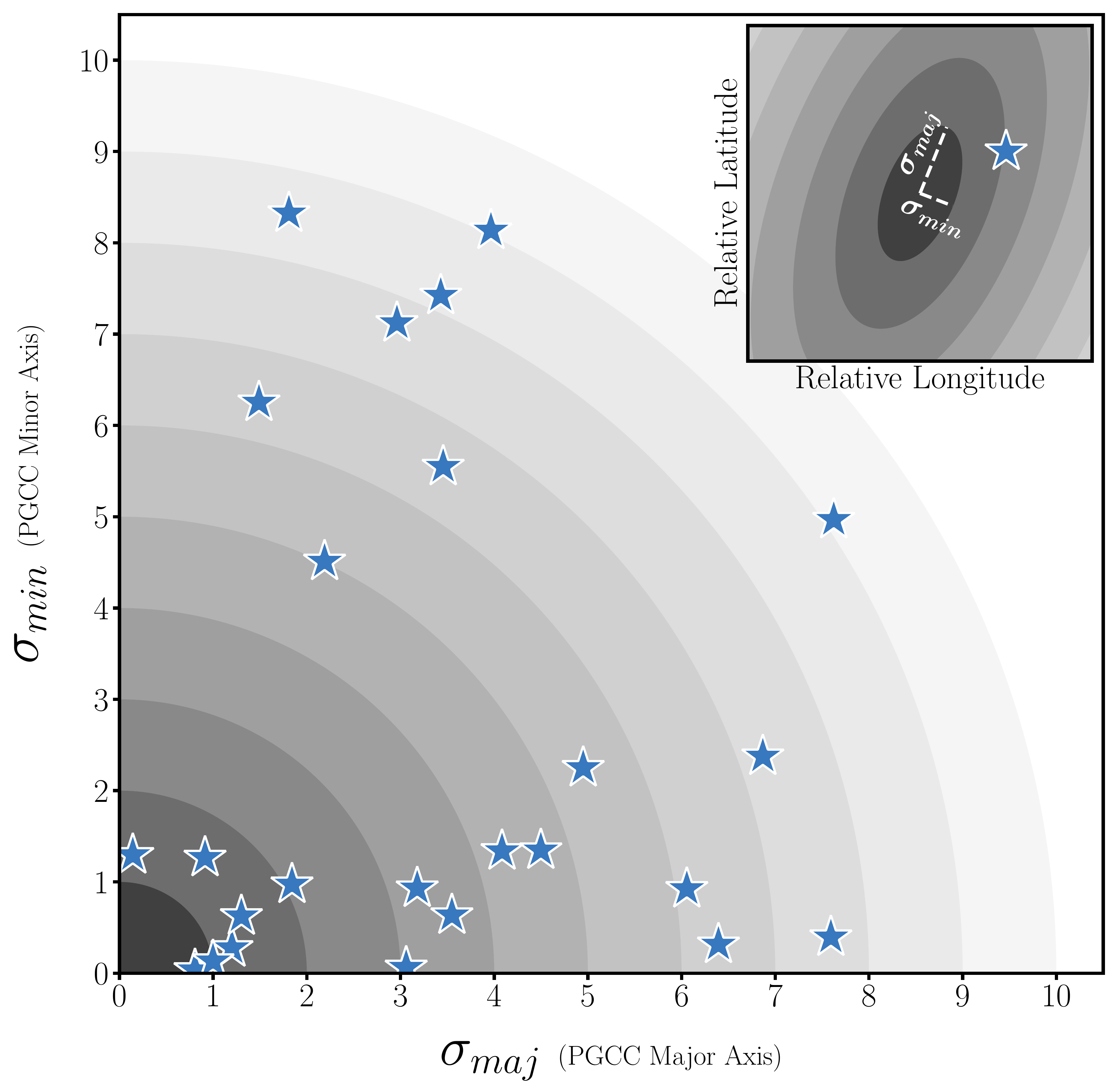}
\caption{Location of the chosen UV absorption sightlines relative to each of their nearby PGCC sources. The axes are in units of $\sigma_{\mathrm{maj}}$ and $\sigma_{\mathrm{min}}$, the major and minor axes of the 2D elliptical Gaussian fits to each PGCC source as listed in \citet{PlanckCollaboration2015}, as shown in the inset image.}
\label{bullseye}
\end{figure}

\subsection{Data Reduction}

All spectra of the identified sightlines in this work were taken using the high-resolution, far-UV E140H echelle mode of the STIS instrument. Spectra were reduced using \verb|Python| routines to co-add individual exposures, continuum fit, and normalize the data in the vicinity of the spectral lines of interest. The resulting normalized spectra were then used as input in the \verb|Fortran| line-profile fitting program FITS6P \citep{Welty1994}. For each sightline, we analyzed the combined spectrum in the vicinity of $^{12}$CO, C\tsc{i} (and its fine-structure excited states C\tsc{i}$^*$ and C\tsc{i}$^{**}$), and O\tsc{i} absorption lines. The specific spectral line analyzed for each of these species is dependent on the spectrum's signal-to-noise ratio in the vicinity of the line, as well as the spectrum's wavelength coverage. Each STIS exposure captures a $\sim$200$\,\angstrom$ region of the entire 1140--1700$\,\angstrom$ STIS far-UV domain. As such, the specific spectral lines available for analysis are not uniform across our entire dataset (for example, some sightlines' spectral coverage contains the C\tsc{i} $\lambda$1270 multiplet, while others instead contain the $\lambda$1328 multiplet). 

\begin{figure}[!t]
\centering
\includegraphics[scale=0.36]{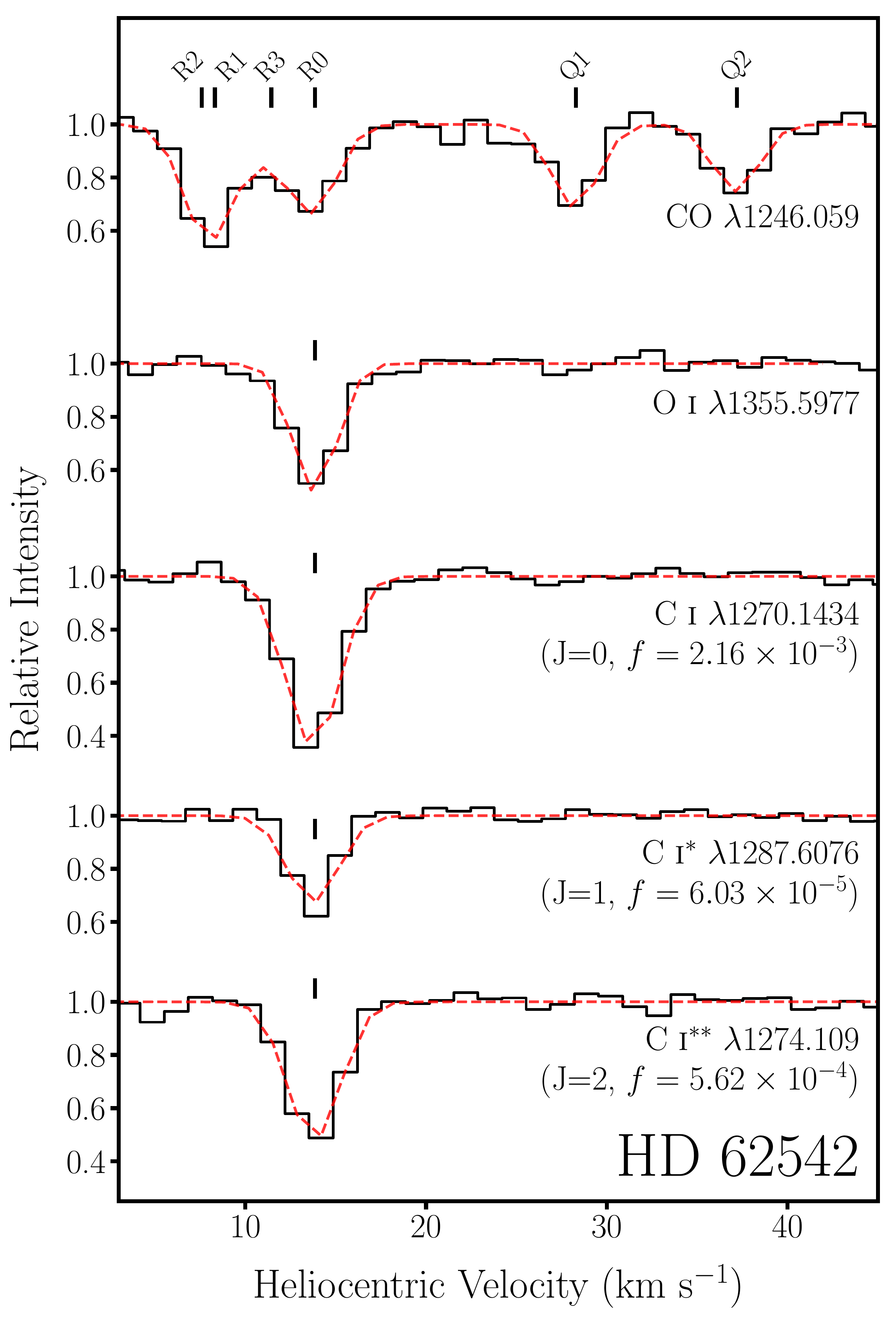}
\caption{Selected regions of the \emph{HST} E140H far-UV spectrum toward HD~62542. The CO molecular absorption band near $\lambda$1246.059, as well as the O\tsc{i}~$\lambda$1355.5977, C\tsc{i}~(J=1)~$\lambda$1287.6076, and C\tsc{i}~(J=2)~$\lambda$1274.109 atomic line regions are shown from top to bottom, respectively. The horizontal axis represents the heliocentric radial velocity relative to each respective rest wavelength, and the vertical axis represents the intensity relative to the continuum flux used to normalize each region. Normalized spectral data are shown as black histograms, with the results of our FITS6P line-profile fitting over-plotted as red dashed lines. This spectrum contains a single absorption component near $v~=~14~\mathrm{km~s^{-1}}$ for all investigated lines. The central velocities of the individual lines used to generate the profile are denoted by black ticks marks. For the CO absorption band, the individual rovibrational lines that comprise the band are annotated, with the R0 line representing the central velocity.}
\label{hd62542}
\end{figure}

\begin{figure}[!t]
\centering
\includegraphics[scale=0.36]{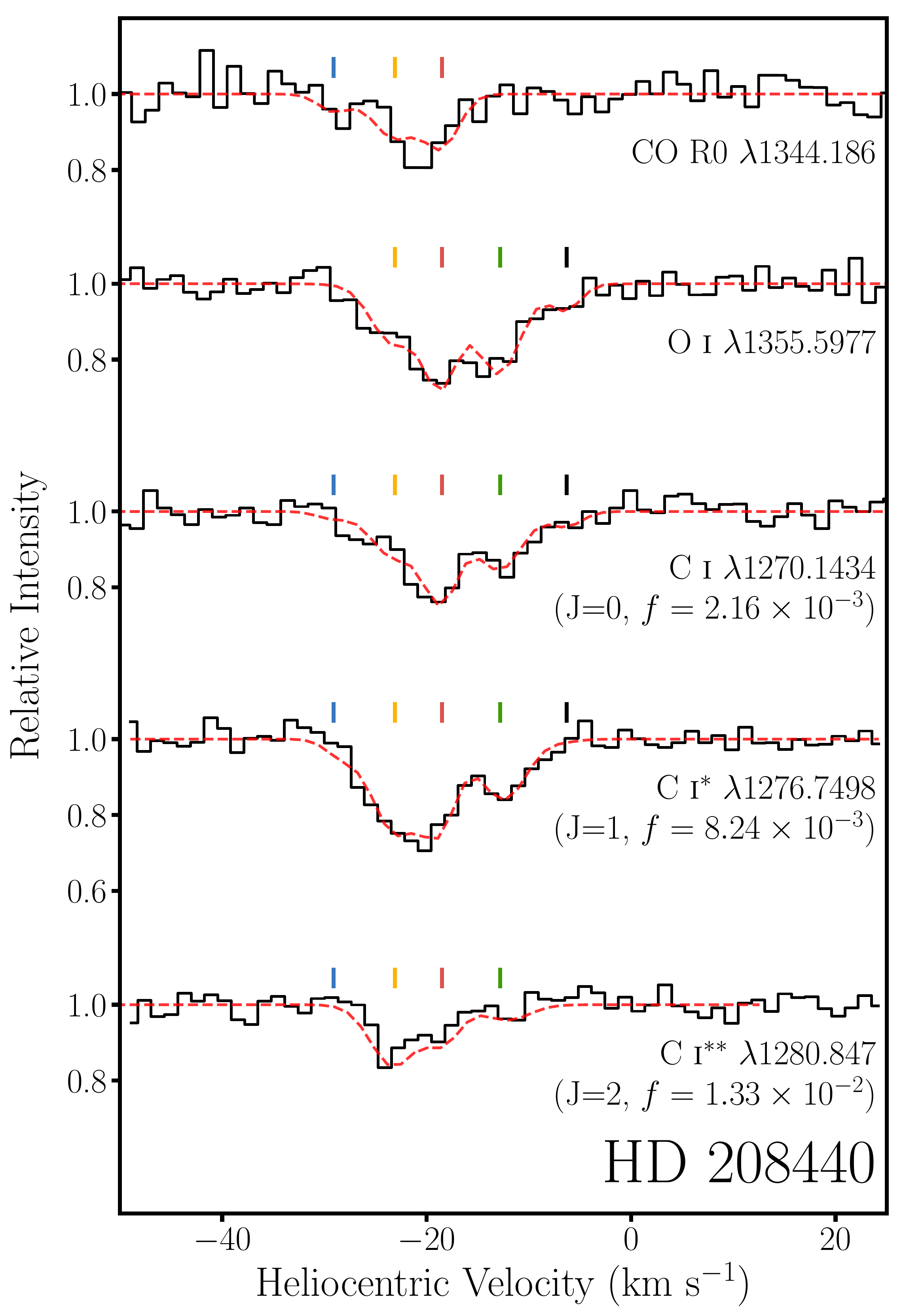}
\caption{Selected regions of the spectrum toward HD~208440, similar to Figure \ref{hd62542}. The CO molecular absorption band near $\lambda$1344.186, as well as the O\tsc{i}~$\lambda$1355.5977, C\tsc{i}~(J=0)~$\lambda$1270.1434, C\tsc{i}~(J=1)~$\lambda$1279.8904, and C\tsc{i}~(J=2)~$\lambda$1280.847 atomic line regions are shown from top to bottom, respectively. Here, the S/N ratio is lower than in Figure \ref{hd62542}, and the absorption profiles are defined by multiple velocity components. The separate components present in each profile are denoted by different colored tick marks. Due to the relative weakness of the CO absorption, only the R0 rovibrational line is present in this spectrum.}
\label{hd208440}
\end{figure}

In order to analyze the gas within the PGCC source, we build cloud-component models for each sightline's spectrum, which allows us to identify and isolate components specifically arising from the PGCC gas. This method lies in contrast to other ISM absorption-line analysis methods which integrate over the sightline's entire absorption profile and thus entangle disparate, independent regions of the ISM. For each absorption component in our models, we determine the column density \emph{N}, line width parameter \emph{b}, and central velocity \emph{v}. We isolate the component that we deem to be associated with the PGCC source by identifying the component with the largest CO column density. While many sightlines (12/25) contain multiple components of CO absorption, the total CO column density is typically dominated by one component. The central velocity of each sightline's selected absorption component is listed in Table \ref{results_table}.

In order to determine the total hydrogen column density, N(H$_{\mathrm{tot}}$), we utilize the results of \citet{Meyer1998}, which utilized \emph{HST} Goddard High Resolution Spectrograph (GHRS) observations of 13 sightlines to compare the strength of the O\tsc{i} $\lambda$1355.598 spectral line to previous measurements of the abundance of H\tsc{i} and H$_2$ along the same sightlines. They found a remarkably consistent relationship between the abundance of interstellar oxygen and the total hydrogen column density of 10$^6$ O/H = 319 $\pm$ 14. Using this, we convert our measured O\tsc{i} column densities to total hydrogen column densities, allowing us to determine the molecular hydrogen fraction, $f_{H_2} = 2*\mathrm{N(H_2)/N(H_{tot})}$.

Figures \ref{hd62542} and \ref{hd208440} contain example spectra for two sightlines, along with our FITS6P profile fitting results. Shown are data near CO and O\tsc{i} lines of interest, as well as individual fine-structure states of C\tsc{i}, which are fit simultaneously using common \emph{b} and \emph{v} parameters and then summed to determine the total C\tsc{i} column density. To aid in comparison between the two figures, the oscillator strength \emph{f} is listed for each C\tsc{i} line. These particular sightlines were chosen to illustrate the variety of spectra present in this dataset. HD~62542 (Figure \ref{hd62542}) is a high signal-to-noise spectrum whose profile is defined by a single component with a large quantity of gas with significant molecular content (as indicated by the strong O\tsc{i} and CO absorption, respectively). Conversely, HD~208440 (Figure \ref{hd208440}) has a lower signal-to-noise ratio, lower overall gas content, relatively little molecular content, and has a profile defined by multiple absorption components.

\subsection{Determining H$_2$ Column Densities}
While the 1140--1700$\,\angstrom$ STIS far-UV wavelength regime covers a wealth of metal lines and CO absorption bands, it lacks any direct H$_2$ absorption features, necessitating the use of indirect methods and/or previously published values to determine the H$_2$ column density. Further complicating matters, our study intends to focus on individual components of the absorption profile, whereas many prior studies were only able to determine N(H$_2$) for the entire sightline, due to the lack of resolution afforded by previous missions such as \emph{FUSE}.

\begin{figure}[!ht]
\centering
\includegraphics[scale=0.24]{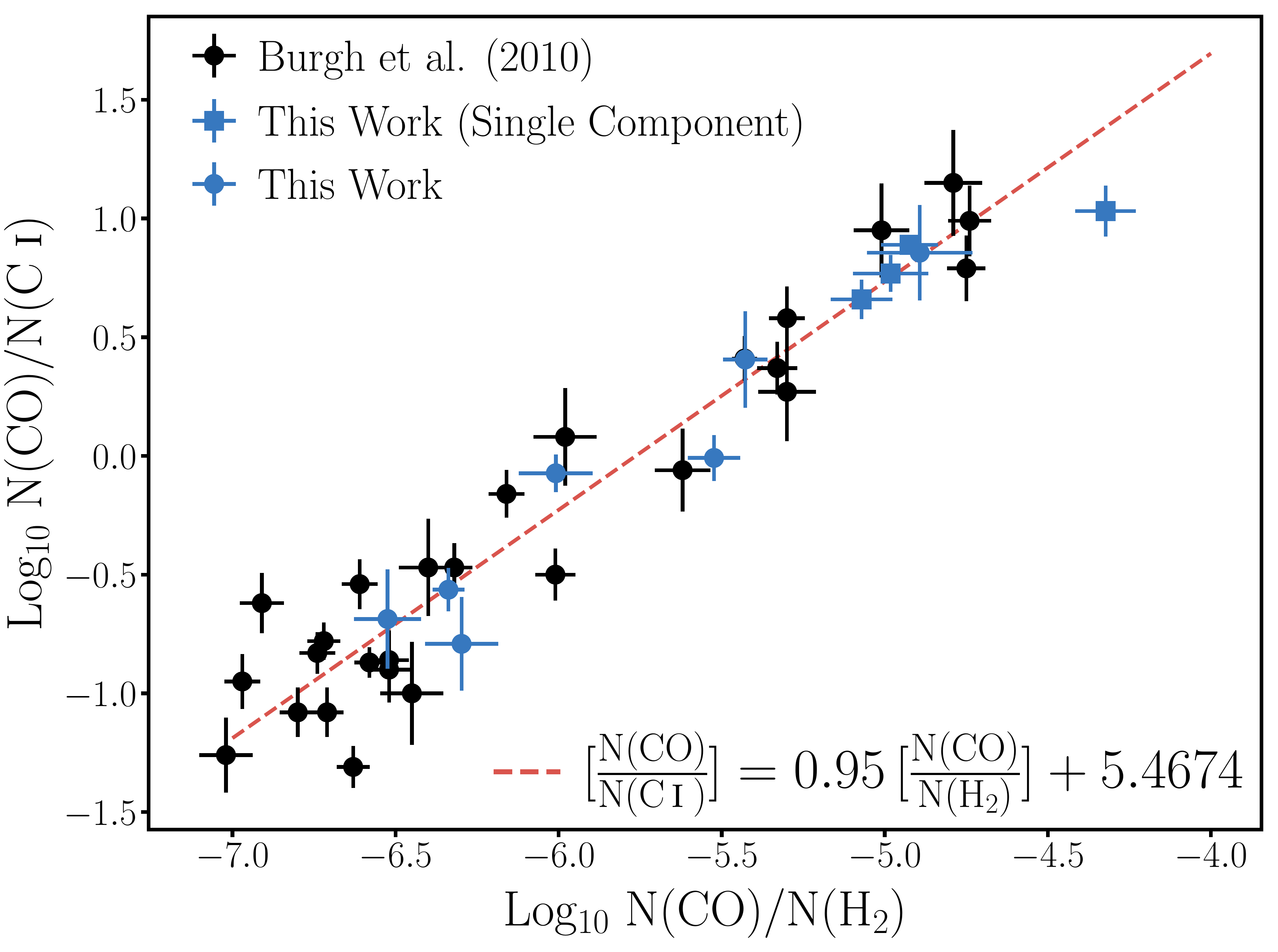}
\caption{Empirical relationship between the logarithms of the CO/C\tsc{i} and CO/H$_2$ ratios for a total of 35 absorption sightlines. Black points are values taken from \citet{Burgh2010}. Blue points represent the combination of our measurements of N(CO) and N(C\tsc{i}) with previously published N(H$_2$) values from the literature. Sightlines whose absorption profiles are defined by a single component are noted (see text for discussion). The red dashed line represents an empirical fit to the combined dataset.}
\label{h2_method}
\end{figure}

To remedy this, we utilize results from \citet{Burgh2010} which found a strong correlation between the CO/C\tsc{i} and CO/H$_2$ ratios. A similar correlation is also seen when our N(CO) and N(C\tsc{i}) measurements are combined with previously published N(H$_2$) values compiled in \citet{Gudennavar2012}, which contains sightline-integrated values for 12 of the sightlines in our sample. Using the combined datasets, we derive an empirical relationship between the two ratios (Figure \ref{h2_method}):
\begin{equation} \label{h2_equation}
\mathrm{Log_{10}\; \frac{N(CO)}{N(C\tsc{i)}}} = 0.95\ast\mathrm{Log_{10}\; \frac{N(CO)}{N(H_2)}} + 5.4674
\end{equation}

With this, we can estimate N(H$_2$) indirectly using direct measurements of N(CO) and N(C~\textsc{i}). While this relationship was derived using sightline-integrated column densities, we assume for this work that it also holds when analyzing the N(CO) and N(C\tsc{i}) values for an individual component. Ideally, in order to understand if this relationship is similar for single-component and sightline-integrated measurements, we would perform an independent analysis on sightlines whose profiles contain a single absorption component across all of the species of interest and compare to a separate sample of sightlines with multiple components. However, of the 12 sightlines within our sample with previously published N(H$_2$) values, only four (HD~24398, HD~24534, HD~62542, and HD~203532) have spectral profiles defined by a single component, limiting our ability to perform a robust analysis of the effects of integrating the entire profile versus isolating single components. Sightlines with single components have been noted in Figure \ref{h2_method} for clarity.

To check the accuracy of this method, we apply this empirical relationship to the 12 sightlines which have previously measured N(H$_2$) values and compare our results with these previously published quantities. We find that in all cases, this method correctly predicts N(H$_2$) to within 0.35 dex (i.e. a factor of $\sim$2).

\section{Results}

\capstartfalse
\renewcommand{\arraystretch}{1.3}
\setlength{\tabcolsep}{6pt}
\begin{deluxetable*}{llrrcccccll}[]
\tablewidth{7in}
\tablecaption{Sightline Parameters and Column Density Results}
\tablehead{\colhead{\phm{\hspace{-8pt}}HD \#} & \colhead{\phm{\hspace{-1pt}}PGCC Source} 
& \colhead{Sep.$^{\mathrm{a}}$}\phm{\hspace{-10pt}} & \colhead{$v^\mathrm{b}$}
& log N(CO) & \emph{b$_{\mathrm{CO}}{}^{\mathrm{c}}$\phm{\hspace{-5pt}}}
& log N(C\tsc{i})$^\mathrm{d}$ & \emph{b$_{\mathrm{C\tsc{i}}}{}^{\mathrm{c}}$\phm{\hspace{-5pt}}}
& log N(H$_{\mathrm{tot}}$)$^\mathrm{e}$
& log N(H$_2$)$_{\mathrm{UV}}$$^\mathrm{f}$
& log N(H$_2$)$_{\mathrm{PGCC}}$$^\mathrm{g}$\phm{\hspace{-8pt}}}
\startdata
\multicolumn{2}{l}{\phm{\hspace{-6pt}}CO-bright}&\\
\hline
   24398 & 162.90-16.59  & 9.1 &  14.4 & 15.69 $\pm$  0.08 & 0.8 & 14.92 $\pm$  0.01 & 1.4 &21.27 $\pm$  0.05 & 20.67 $\pm$  0.11$^{1}$ & 20.49 $\pm$  0.12\\
   24534 & 162.84-17.15  & 3.1 &  15.2 & 16.02 $\pm$  0.03 & 0.8 & 15.13 $\pm$  0.03 & 2.0 &21.37 $\pm$  0.05 & 20.84 $\pm$  0.17 & 20.46 $\pm$  0.19\\
   62542 & 255.83-9.21   & 1.3 &  13.8 & 16.49 $\pm$  0.08 & 0.5 & 15.46 $\pm$  0.11 & 1.0 &21.44 $\pm$  0.15 & 20.81 $\pm$  0.27$^{2}$ & 20.47 $\pm$  0.47\\
  108927 & 302.11-14.92  & 8.2 &  10.2 & 15.30 $\pm$  0.21 & 0.6 & 14.80 $\pm$  0.01 & 2.2 &21.14 $\pm$  0.11 & 20.53 $\pm$  0.15 & 20.53 $\pm$  0.20\\
  147683 & 345.01+10.04  & 2.1 &  -0.9 & 15.79 $\pm$  0.20 & 0.6 & 14.88 $\pm$  0.03 & 1.3 &21.29 $\pm$  0.09 & 20.58 $\pm$  0.14 & 20.32 $\pm$  0.30\\
  203532 & 309.83-31.87  & 4.7 &  15.0 & 15.57 $\pm$  0.08 & 1.0 & 14.91 $\pm$  0.02 & 2.3 &21.32 $\pm$  0.06 & 20.64 $\pm$  0.09$^{1}$ & 20.11 $\pm$  0.19\\
  210839 & 103.46+2.81   & 9.1 & -14.8 & 15.38 $\pm$  0.04 & 1.2 & 14.82 $\pm$  0.01 & 1.8 &21.38 $\pm$  0.06 & 20.55 $\pm$  0.18 & 21.03 $\pm$  0.43\\
  254755 & 189.10+2.96   & 5.4 &   6.4 & 15.82 $\pm$  0.24 & 0.5 & 15.11 $\pm$  0.20 & 2.1 &21.79 $\pm$  0.09 & 20.83 $\pm$  0.14 & 21.31 $\pm$  0.46\\
&&&&&&&&\\
\multicolumn{2}{l}{\phm{\hspace{-6pt}}CO-dark} & \\
\hline
    4768 & 122.79-3.24   & 0.8 & -18.5 & 14.67 $\pm$  0.10 & 1.7 & 14.38 $\pm$  0.33 & 1.7 &20.84 $\pm$  0.22& 20.13 $\pm$  0.18 & 20.08 $\pm$  0.28\\
   13841 & 134.21-4.20   & 3.6 & -13.7 & 14.36 $\pm$  0.12 & 0.8 & 14.20 $\pm$  0.02 & 1.3 &20.87 $\pm$  0.23 & 19.95 $\pm$  0.18 & 20.48 $\pm$  0.14\\
   23180 & 160.46-17.99  & 6.4 &  14.7 & 14.64 $\pm$  0.33 & 1.0 & 14.78 $\pm$  0.03 & 1.4 &21.12 $\pm$  0.05& 20.55 $\pm$  0.14 & 21.97 $\pm$  0.16\\
   23478 & 160.52-17.04  & 6.1 &  11.0 & 14.87 $\pm$  0.07 & 0.8 & 14.82 $\pm$  0.11 & 1.2 &20.93 $\pm$  0.14& 20.57 $\pm$  0.19 & 21.31 $\pm$  0.35\\
   25443 & 143.67+6.92   & 6.4 &   4.2 & 14.31 $\pm$  0.10 & 1.1 & 14.21 $\pm$  0.22 & 2.0 &21.04 $\pm$  0.13& 19.96 $\pm$  0.19 & 20.34 $\pm$  0.22\\
   43582 & 189.10+2.96   & 3.3 &   7.3 & 13.93 $\pm$  0.10 & 1.0 & 14.65 $\pm$  0.03 & 2.4 &21.65 $\pm$  0.06 & 20.45 $\pm$  0.21 & 21.31 $\pm$  0.46\\
   72350 & 262.63-3.24   & 1.6 &  20.2 & 14.05 $\pm$  0.08 & 1.5 & 14.19 $\pm$  0.05 & 2.0 &20.84 $\pm$  0.12 & 19.96 $\pm$  0.20 & 21.02 $\pm$  0.17\\
  108002 & 300.00-2.77   & 8.5 &   9.9 & 13.63 $\pm$  0.08 & 1.0 & 13.95 $\pm$  0.22 & 1.3 &20.64 $\pm$  0.25 & 19.72 $\pm$  0.20 & 20.81 $\pm$  0.25\\
  112999 & 304.27+1.69   & 7.7 &   8.1 & 12.85 $\pm$  0.99 & 1.5 & 13.96 $\pm$  0.12 & 2.0 &21.04 $\pm$  0.15 & 19.77 $\pm$  0.07 & 21.17 $\pm$  0.44\\
  124314 & 312.84-0.71   & 4.3 &   2.8 & 14.01 $\pm$  0.09 & 1.0 & 14.36 $\pm$  0.05 & 1.0 &21.16 $\pm$  0.10& 20.13 $\pm$  0.20 & 21.73 $\pm$  0.22\\
  148594 & 350.77+14.50  & 7.6 &  -4.7 & 12.75 $\pm$  1.28 & 1.0 & 14.23 $\pm$  0.05 & 2.0 &21.31 $\pm$  0.09& 20.06 $\pm$  0.56 & 20.93 $\pm$  0.28\\
  165918 & 10.69-0.13    & 5.0 &  -6.9 & 13.32 $\pm$  0.08 & 1.3 & 14.27 $\pm$  0.02 & 2.3 &21.20 $\pm$  0.07& 20.08 $\pm$  0.23 & 22.18 $\pm$  0.37\\
  185418 & 54.02-2.38    & 7.3 & -11.1 & 14.60 $\pm$  0.06 & 0.7 & 14.55 $\pm$  0.07 & 1.3 &20.93 $\pm$  0.11& 20.30 $\pm$  0.19 & 21.09 $\pm$  0.10\\
  208440 & 104.08+6.36   & 1.2 & -23.1 & 13.65 $\pm$  0.23 & 1.7 & 14.27 $\pm$  0.10 & 2.5 &20.83 $\pm$  0.14 & 20.06 $\pm$  0.19 & 20.37 $\pm$  0.45\\
  208947 & 106.53+9.30   & 6.5 & -14.4 & 14.10 $\pm$  0.08 & 0.5 & 14.84 $\pm$  0.09 & 0.8 &21.04 $\pm$  0.09& 20.63 $\pm$  0.22 & 20.35 $\pm$  0.38\\
  220058 & 110.26-4.87   & 1.0 &  -7.5 & 13.52 $\pm$  0.12 & 2.0 & 14.50 $\pm$  0.10 & 2.8 &21.17 $\pm$  0.10& 20.31 $\pm$  0.22 & 20.40 $\pm$  0.17\\
  232522 & 130.69-6.79   & 1.4 & -17.0 & 13.45 $\pm$  0.07 & 1.2 & 13.92 $\pm$  0.05 & 2.0 &20.85 $\pm$  0.07 & 19.70 $\pm$  0.21 & 20.56 $\pm$  0.07
\enddata

\tablecomments{Results of the absorption line-profile fitting for the single component from each sightline identified to be associated with the PGCC source.\\
$^\mathrm{a}$ Separation on the sky between the stellar sightline and the center of the PGCC source, expressed in units of $\sigma$, which is derived from the PGCC source's elliptical Gaussian FWHM parameters (see Figure \ref{bullseye}).\\
$^\mathrm{b}$ Central velocity in km~s$^{-1}$ of the absorption component associated with the PGCC source (i.e. that with the largest N(CO)).\\
$^\mathrm{c}$ Line-width parameters used in the profile fitting of the CO and C\tsc{i} species, respectively.\\
$^\mathrm{d}$ Summation of the column density of the three C\tsc{i} fine-structure states (J=0,1,2).\\
$^\mathrm{e}$ Total hydrogen column density determined indirectly from the O\tsc{i} column density (see \citet{Meyer1998} for details).\\
$^\mathrm{f}$ When noted, these values and their associated errors are taken from the following references: (1) \citet{Cartledge2001}; (2) \citet{Jensen2005}. In all other cases, N(H$_2$) is calculated using the method described in \S2.3, with the errors representing the combination of measurement uncertainties in the CO and C\tsc{i} column densities and the uncertainties in the derived empirical relationship. \\
$^\mathrm{g}$ Molecular hydrogen column density listed for the source (defined as the $1\sigma$ region, in the notation of Figure \ref{bullseye}) in the PGCC catalogue.}
\label{results_table}
\end{deluxetable*}
\renewcommand{\arraystretch}{1.0}
\setlength{\tabcolsep}{6pt}
\capstarttrue

The results of our absorption-line profile fitting are shown in Table \ref{results_table}. The selected targets span a broad range in total sightline column density ($6.9 \times 10^{20} < \mathrm{N(H_{tot})} < 1.1 \times 10^{22}$ cm$^{-2}$) and sightline molecular fraction ($0.1 < f_{H_2} < 0.75$). CO is detected toward all of our investigated targets, albeit often in quantities below the detection limit of radio surveys. Recent photodissociation region simulations by \citet{Gong2017} predict that CO should be present in quantities invisible to radio emission but detectable by UV absorption (i.e. N(CO)~$\sim$~10$^{13}$~cm$^{-2}$) down to total hydrogen column densities on the order of a few times 10$^{20}$~cm$^{-2}$ (with the exact value dependent on the volume density of the gas), which is  consistent with the values seen here. 

In order to compare to the aforementioned radio surveys, we define two classes of sightlines within our sample --- those that are ``CO-bright'' (i.e. have N(CO)~$>$~10$^{15}$~cm$^{-2}$ and would be detectable via traditional radio emission observations) and, conversely, those that are ``CO-dark.'' Only 8/25 (32\%) of these sightlines are CO-bright, a rate comparable to the $\sim$29\% detection rate of \citet[Figure 3]{Liu2018}. That study sought to probe a sample representative of the entire PGCC catalogue, yet was nonetheless biased toward higher column density sources in an effort to ensure detection via radio follow-up. (\citet{Liu2018} does not include the specific sample of PGCC sources they chose for follow-up, and thus their specific source distribution \textbf{is} not included in Figure \ref{prev_pgcc_studies}. For a qualitative comparison, their N(H$_2$) distribution (seen in their Figure 2) peaks at column density of ~$\sim$10$^{21}$~cm$^{-2}$, whereas the distribution of N(H$_2$) values for the entire PGCC sample peaks around 3$\times 10^{20}$~cm$^{-2}$).

\begin{figure}[]
\centering
\includegraphics[scale=0.22]{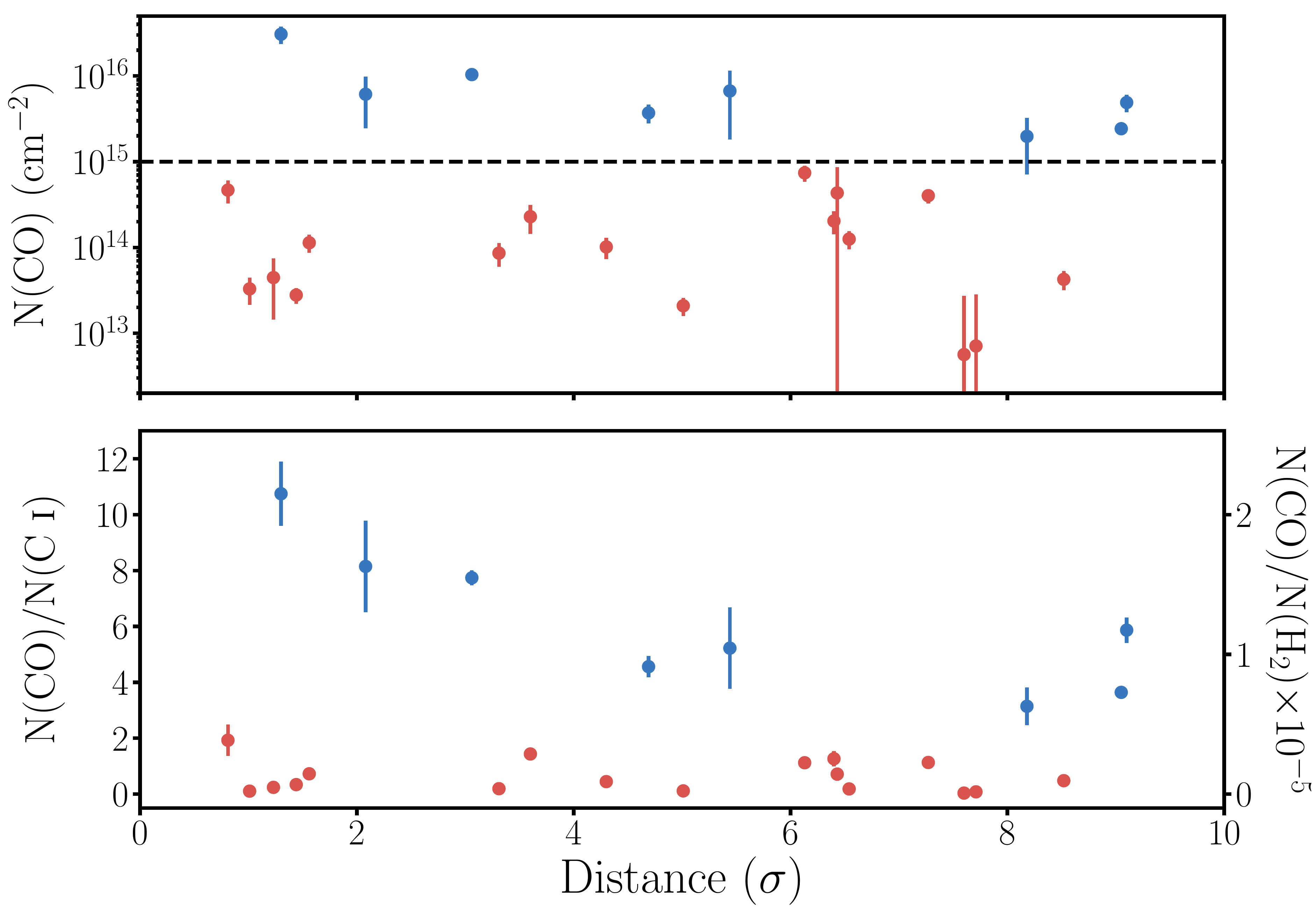}
\caption{Dependence of N(CO) (top panel) and N(CO)/N(C\tsc{i}) on projected distance away from the center of the nearby PGCC source. The x-axis is scaled by units of $\sigma$, which combines the two-dimensional (major and minor axis) FWHM parameters of the PGCC source. For the column density measurements, we have isolated the component along each sightline with the largest CO column density. This component is assumed to be associated with the PGCC source. Points with N(CO) $>$ 10$^{15}$~cm$^{-2}$ are coded blue, while points with N(CO) $<$ 10$^{15}$~cm$^{-2}$ are coded red.}
\label{distance_dependence}
\end{figure}

Using this distinction, we see that CO-bright sources exhibit a unique behavior in their N(CO) values and CO/C\tsc{i} ratio (and corresponding CO/H$_2$ ratio, as per the empirical relationship discussed in \S2.3). These properties appear to vary spatially with projected distance away from the center of the PGCC source, decreasing as the sightlines probe regions further from the central core of the PGCC source. Conversely, CO-dark sources show no apparent correlation between these properties and proximity to the PGCC source, instead remaining steady at N(CO) and CO/C\tsc{i} values consistently below what is seen in the CO-bright sample (Figure \ref{distance_dependence}). The CO/C\tsc{i} ratio is relevant here due to the results of \citet{Burgh2010}, which found that the CO/C\tsc{i} and CO/H$_2$ ratios were correlated (as discussed in \S2.3 and Figure \ref{h2_method}), and served as useful indicators of how translucent a particular sightline was. That study determined that values of CO/H$_2$ = 10$^{-6}$ and CO/C\tsc{i} = 1 served as good cutoffs between ``diffuse'' and ``translucent'' sightlines. We note that, for our study, a CO/C\tsc{i} ratio of $\sim$2 serves as an equivalent discriminator between our CO-bright and CO-dark sightlines (i.e. all of our PGCC components with N(CO)~$>$~10$^{15}$~cm$^{-2}$ also have CO/C\tsc{i}~$>$~2, and vice-versa). 

\begin{figure}[!h]
\centering
\includegraphics[scale=0.24]{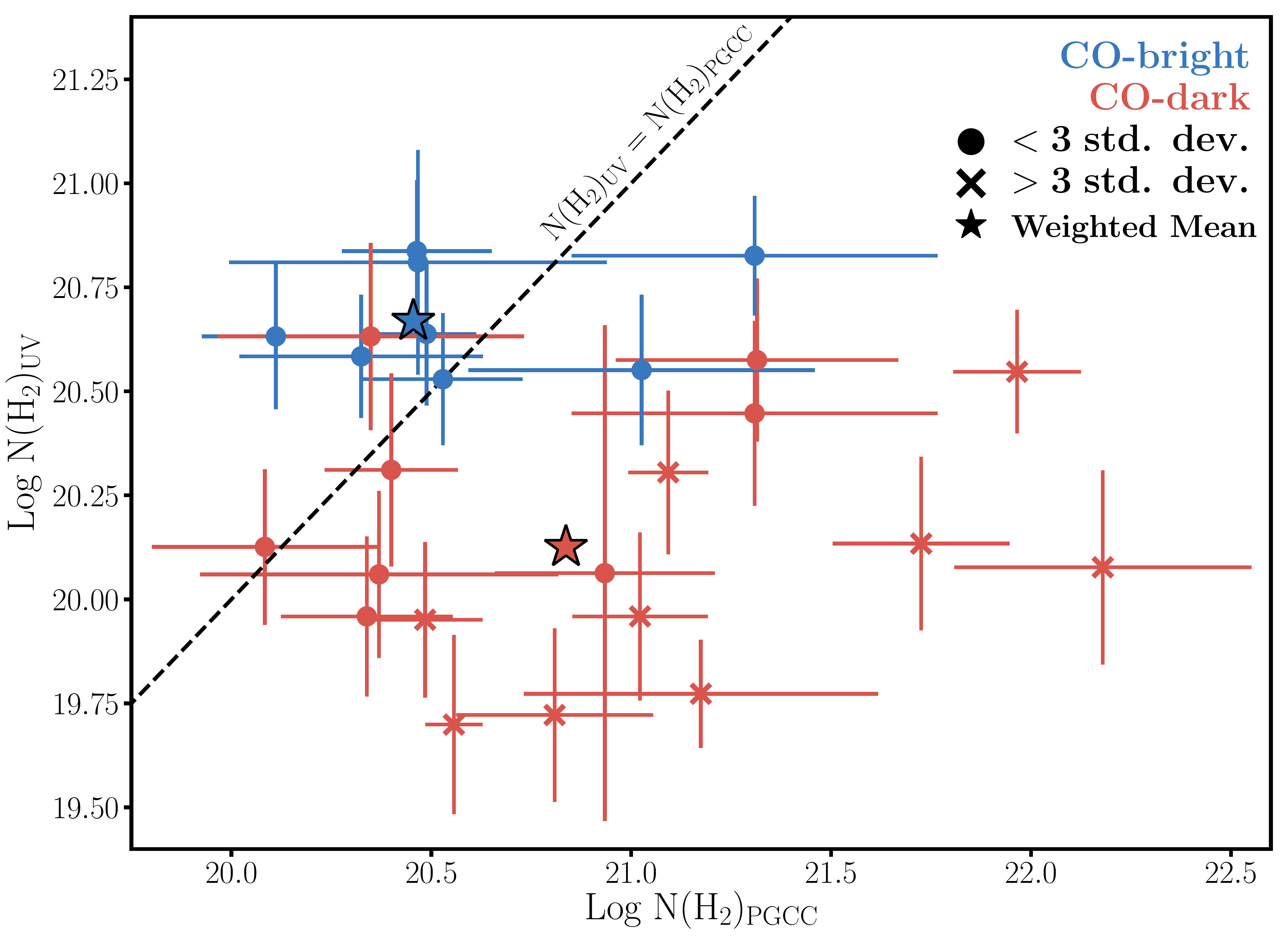}
\caption{Comparison of the H$_2$ column densities reported in the PGCC catalogue for each source to the value measured from UV absorption. Each point's color corresponds to its CO-bright (blue) or CO-dark (red) designation. Each point's symbol corresponds to the discrepancy between the two measurements, with circle symbols denoting values that lie within three standard deviations of each other, and ``x'' symbols denoting those whose measurements are discrepant by more than three standard deviations. The weighted mean of both the CO-bright and CO-dark samples is indicated as well. Of note is the fact that all CO-bright points lie within three std. devs., while a significant number of CO-dark points have measurements which are discrepant.}
\label{h2_comparison}
\end{figure}

Finally, we compare the \nh{PGCC} values for each source as given in the PGCC catalogue to those derived from UV measurements of CO and C\tsc{i} using the relationship described in \S2.3. These values are listed in the two right-most columns in Table \ref{results_table}, and a comparison between these values is shown in Figure \ref{h2_comparison}. Among our CO-bright sightlines, all measurements of \nh{UV} and \nh{PGCC} agree to within 3 standard deviations (where 1 standard deviation represents the combined errors in both \nh{UV} and \nh{PGCC}). Conversely, among our CO-dark sightlines, nearly half (8/17) have \nh{} measurements that differ by more than 3 standard deviations. In all 8 of these cases, \nh{PGCC} is larger than \nh{UV}. We note than in both the CO-bright and -dark samples, the sightlines cover a wide range of angular separations, from very near to the source ($<2\sigma$), to very far away ($>8\sigma$). This confirms that this effect is not simply due to the relative separation between the source and the sightline, but indicative of the presence of more complex structure giving rise to these two classes of sightlines.

\section{Discussion}
In order to understand what gives rise to the observed distinctions between CO-bright and CO-dark sightlines, we first aim to understand the discrepancies between \nh{} values described above. In understanding these discrepancies, it is important to discuss the nature of each type of observation. As described in \citet{PlanckCollaboration2011}, N(H$_2$)$_{\mathrm{PGCC}}$ for each source is determined by measuring the dust emissivity and subtracting off the warm background component as seen by the InfraRed Astronomical Satellite (IRAS), in order to isolate the contribution specifically arising from the cold dust within the source. Using this dust emissivity, they fit a blackbody emission law of the form:
\begin{equation}
S_{\nu} = \Omega_c \: \kappa_{\nu} \: B_{\nu}(T) \: N(H_2) \: \mu \: m_H
\end{equation}
where $\Omega_c$ represents the sold angle covered by the source ($\Omega_c=\pi\:\sigma_{maj}\;\sigma_{min}$, see inset of Figure \ref{bullseye}), $\kappa_{\nu}$ is the dust opacity as a function of $\nu$, $\kappa_{\nu} = 0.1(\nu/1\:\mathrm{THz})^2$~cm$^{2}$~g$^{-1}$ \citep{Beckwith1990}, $B_{\nu}$(T) is the Planck blackbody function, and $\mu$ and $m_H$ represent the mean molecular weight of H$_2$ and mass of atomic hydrogen, respectively. For PGCC sources, the dust emissivity map from the \emph{Planck} 857$\;$GHz band was used as an input, and the resulting N(H$_2$) was then integrated over the 1$\sigma$ region of the detected source to derive the \nh{PGCC} values listed in the catalogue. Due to the width of the \emph{Planck} beam and the nature of the source detection algorithm which requires the input dust maps and the background IRAS maps to be convolved to the same 5 arcminute FWHM resolution (corresponding to physical sizes of $\sim$0.15 pc for nearby ($\sim$100~pc) sources or $\sim$15 pc for distance ($\sim$10~kpc) sources), PGCC sources have $\sigma_{maj},\sigma_{min}$ values that are typically of order $\sim$5--10 arcmin. The corresponding mean \nh{PGCC} values therefore represent averages over dozens to hundreds of square arcminutes on the sky. In contrast, an absorption sightline is only sensitive to the ISM inside a single ``beam'' of $\sim$ 10$^{-4}$ arcseconds through the region. Because these two methods are sensitive to such considerably different scales, any discrepancies between these two measurements may yield insight into the \nh{} structure within the ISM.

The prevalence of CO-dark sightlines with \nh{UV}~$<<$~\nh{PGCC} (Figure \ref{h2_comparison}) at a broad range of angular separations indicates that the \nh{} distribution in the environments surrounding the PGCC sources has a ``patchy'' structure, with some fraction of the gas being relatively devoid of molecular material. The presence of the patchiness would serve to increase the surface area subject to the interstellar UV radiation field, and thus prevent any coherent molecular regions from forming. The detailed structure within these regions would be unresolved by \emph{Planck} observations due to a large beam width, but the narrow beam of an absorption sightline has some chance of traveling through relatively little molecular material. Conversely, the agreement between \nh{UV} and \nh{PGCC} in all of our CO-bright sightlines is interpreted as evidence that the PGCC sources associated with these sightlines are embedded in smoother envelopes of translucent gas, with properties akin to the translucent regions discussed in \citet{Burgh2010}. This would explain the overall higher abundance of CO along these sightlines, as these structures would have the bulk of their material shielded from UV radiation that is capable of dissociating H$_2$ and CO. Additionally, the spatial dependence observed in both CO and the CO/C\tsc{i} in Figure \ref{distance_dependence} would be expected from a more traditional cloud-like structure, with a central, denser region tapering off into more diffuse material at the outskirts of the cloud. 

As detailed in \S3, thresholds of N(CO)$\,\sim\,$10$^{15}$~cm$^{-2}$ and CO/C\tsc{i}$\,\sim\,$2 both appear to distinguish these two classes, indicating that this is when the ISM gas becomes sufficiently self-shielded from UV radiation to allow larger, more translucent molecular regions to form. While this threshold is slightly higher compared to the value of CO/C\tsc{i}=1 used in \citet{Burgh2010} to distinguish between diffuse and translucent sightlines, we note that that study used column density values that were derived by integrating over the entire absorption profile, whereas this work isolates and studies individual components. Since neutral carbon is, in general, more abundant in the ISM than CO, the effect of integration over the entire profile would be to drive this ratio to smaller values than what would be expected from isolating a single component of CO-bearing gas. Even with this slight caveat, it is clear that the \citet{Burgh2010} definition of translucence corresponds neatly with our definition of CO-bright.

\section{Conclusions}
We have identified a sample of 25 UV absorption-line systems whose sightlines probe regions of cold dust emission identified by \citet{PlanckCollaboration2015}, representing the first time that a sample of PGCC sources has been studied in the UV regime. This has allowed us to probe a set of sources whose N(H$_2$) distribution more accurately represents that of the entire PGCC catalogue, in contrast with previous studies based on radio observations which are naturally biased toward sources with large N(H$_2$) values. By building cloud-component models for each sightline's spectrum, we have isolated the absorption component within each spectrum that arises from a nearby PGCC source, rather than using results derived from integrating across the entire absorption profile. This enables us to specifically probe the gas in the local vicinity of these sources, yielding a more accurate depiction of the true physical conditions within these sources.

CO is detected along all sightlines, in quantities similar to those predicted by recent photodissociation models by \citet{Gong2017}. These quantities are typically below the threshold for detection through CO emission in the radio regime, but we find a rate of ``CO-bright'' detections similar to \citet{Liu2018}, the largest radio study of PGCC sources to date. We observe that our CO-bright sources exhibit spatial dependence in their N(CO) values and CO/C\tsc{i} ratios, while our CO-dark sources show no such spatial dependence. Furthermore, for our CO-bright sources, the derived \nh{UV} values match the \nh{PGCC} values listed in \citet{PlanckCollaboration2015}. While the \nh{UV} values for our CO-dark sources are also often similar to the \nh{PGCC} values, there is a significant fraction of CO-dark sources where \nh{UV} is substantially less than \nh{PGCC}.
These results are interpreted as evidence that our CO-bright sightlines are probing PGCC sources with smooth, translucent envelopes, while our CO-dark sightlines are probing PGCC sources embedded within patchier, structured environments. 
This structure would be unresolved in the maps of dust emission used to extract the PGCC sources due to the beam width of the \emph{Planck} instruments, but the narrow beam of an absorption sightline would occasionally observe less molecular material than seen by a corresponding \emph{Planck} observation.

\section*{Acknowledgements}
We are thankful to the referee for providing helpful comments which improved the quality of this paper. All of the ultraviolet data used in the work were obtained from the Mikulski Archive for Space Telescopes (MAST). Support for this work was provided by NASA through grant numbers HST-AR-14292 and HST-GO-15104 from the Space Telescope Science Institute, which is operated by AURA, Inc., under NASA contract NAS 5-26555.

Facilities: \facility{HST}
\bibliographystyle{apj}
\bibliography{bibfile}

\clearpage

\end{document}